\documentstyle[12pt]{article}
\def\be{\begin{eqnarray}}
\def\ee{\end{eqnarray}}
\def\ba{\begin{array}}
\def\ea{\end{array}}
\def\p{\phi}
\def\vp{\varphi}
\def\a{\alpha}
\def\ep{\epsilon}
\def\pa{\partial}

\def\G{{\cal G}}

\def\A{{\cal A}}
\def\X{{\cal X}}

\def\D{^{(5)}}
\begin{document}
\begin{center}
{\Large \bf {
IWP Solutions for Heterotic String \\
\vskip 3mm
in Five Dimensions}}
\end{center}
\vskip 1cm
\begin{center}
{\bf \large  {Alfredo Herrera-Aguilar}}
\end{center}
\begin{center}
Joint Institute for Nuclear Research,\\
Dubna, Moscow Region 141980, Russia.\\
e-mail: alfa@cv.jinr.dubna.su
\end{center}
\vskip 0.5cm
\begin{center}
and
\end{center}
\vskip 0.5cm
\begin{center}
{\bf \large  {Oleg Kechkin}}
\end{center}
\begin{center}
Institute of Nuclear Physics,\\
M.V. Lomonosov Moscow State University, \\
Moscow 119899, Russia, \\
e-mail: kechkin@monet.npi.msu.su
\end{center}
%%%%%%%%%%%%%%%%%%%%%%%%%%%%%%%%%%%%%%%%%%%%%%%%%%%%%%%%%%%%%%%%%%%%%%%%%%%%%
\begin{abstract}
We obtain extremal stationary solutions that generalize the
Israel--Wilson--Perj\'es class for the low-energy limit of heterotic string
theory with $n\geq 3$ $U(1)$ gauge fields toroidally compactified from five
to three dimensions. A dyonic solution is obtained using the matrix Ernst
potential (MEP) formulation and expressed in terms of a single real
$3\times 3$--matrix harmonic function. By studying the asymptotic behaviour
of the field configurations we define the physical charges of the field
system. The extremality condition makes the charges to saturate the
Bogomol'nyi--Prasad--Sommerfield (BPS) bound.
\end{abstract}
%%%%%%%%%%%%%%%%%%%%%%%%%%%%%%%%%%%%%%%%%%%%%%%%%%%%%%%%%%%%%%%%%%%%%%%%%%%%%%
%%%%%%%%%%%%%%%%%%%%%%%%%%%%%%%%%%%%%%%%%%%%%%%%%%%%%%%%%%%%%%%%%%%%%%%%%%%%%%
\newpage
%%%%%%%%%%%%%%%%%%%%%%%%%%%%%%%%%%%%%%%%%%%%%%%%%%%%%%%%%%%%%%%%%%%%%%%%%%%%
\section{Introduction}
In effective low energy theories of gravity derived from superstring theory
Einstein gravity is supplemented by additional fields such as the
Kalb--Ramond, gauge fields, and the scalar dilaton  which couples in a
non--tivial way to other fields. These string gravity models preserve the
long--distance behaviour of the mysterious quantum gravity and in special
(BPS saturated) cases exactly reproduce it
\cite{kir}
.
%^1
The bosonic sector of heterotic string theory compactified to three
dimensions on a torus can be parametrized by the $(d+1)\times(d+1)$ and 
$(d+1)\times n$ $Matrix$ $Ernst$ $Potentials$ $\X$ and $\A$
\cite{hk1}--\cite{hk2}
,
%^{2--3}
where $d+3$ is the original space--time dimension and $n$ is the number of
Abelian vector fields.

In this letter we consider the five--dimensional case and suppose a linear
dependence between the MEP $\X$ and $\A$ following the procedure
indicated in our previous work
\cite{hk2}
.
%^{3}
If $n\geq 3$ this leads to non--trivial field configurations which generalize the
the IWP class of solutions of the Einstein--Maxwell (EM) theory. Note that 
in our approach the number of gauge fields is bounded from below. A similar
ansatz arose in the four dimensional case considered in
\cite{hk2} and \cite{bko}
%Refs. 3 and 4
when the minimal number of gauge fields was equal to two (in the general
case $n\geq d+1$
\cite{hk2}).
%^{3}

Furthermore, it is shown that the physical
charges of the obtained solutions saturate the BPS bound as a
consequence of the extremality condition. Among them we identify rotating
black hole--type solutions with both electric and magnetic charges (dyonic
solutions).

Some classes of five--dimensional BPS solutions with trivial and non--trivial
values of electromagnetic charges were obtained in
\cite{ky}--\cite{cfgk}.
%Refs. 5--7.
%%%%%%%%%%%%%%%%%%%%%%%%%%%%%%%%%%%%%%%%%%%%%%%%%%%%%%%%%%%%%%%%%%%%%%%%%%%%
\section{Matrix Ernst Potentials}
We start from the effective field theory of heterotic
string in five dimensions. The action of this theory reads
\be S\D=\int d\D x \mid
G\D\mid^{\frac{1}{2}}\,e^{-\p\D}(R\D+
\p\D_{;M}\,\p^{(5);M}-
\frac{1}{12}\,H\D_{MNP}\,H^{(5)MNP}-
\frac{1}{4}\,F^{(5)I}_{MN}\,F^{(5)IMN}),
\ee
where
\be &&F^{(5)I}_{MN}=\pa
_MA^{(5)I}_N-\pa _NA^{(5)I}_M,
\nonumber \\ &&H\D_{MNP}=\pa
_MB\D_{NP}-\frac{1}{2}A^{(5)I}_M\,F^{(5)I}_{NP}+ \mbox{\rm cycl. perms. of
M,N,P.}
\nonumber
\ee
Here $G\D_{MN}$ is the $5$-dimensional metric,
$B\D_{MN}$ is the anti--symmetric Kalb-Ramond field, $\p\D$ is the dilaton
and $A^{(5)I}_M$ denotes a set ($I=1,\,2,\,...,n$) of $U(1)$ gauge fields.

After the Kaluza-Klein compactification on a two--torus, one obtains the
following set of three--dimensional fields
\cite{ms}-\cite{s1}
:
%^{8--9}

a) scalar fields
\be
G=(G_{pq} \equiv G\D_{p+3,q+3}), \quad
B=(B_{pq} \equiv B\D_{p+3,q+3}),
\nonumber    \\
A=(A^I_p \equiv A^{(5)I}_{p+3}), \quad
\p=\p\D-\frac{1}{2}{\rm ln\,|det}\,G|,
\ee
where $p,q=1,2$.

b)tensor fields
\be
g_{\mu\nu}=e^{-2\p}\left(G\D_{\mu\nu}-G\D_{p+3,\mu} G\D_{q+3,\nu}G^{pq}
\right), \quad
B_{\mu\nu}=B\D_{\mu\nu}-4B_{pq}A^p_{\mu}A^q_{\nu}-
2\left(A^p_{\mu}A^{p+2}_{\nu}-A^p_{\nu}A^{p+2}_{\mu}\right),
\nonumber
\ee
(following A. Sen
\cite{s1}
%^8
we consider the ansatz when $B_{\mu\nu}=0$).

c)vector fields $A^{(a)}_{\mu}=
\left((A_1)^p_{\mu},(A_2)^{p+2}_{\mu},(A_3)^{4+I}_{\mu}\right)$
($a=1,2,3,4, 4+I$)
\be
&&(A_1)^p_{\mu}=\frac{1}{2}G^{pq}G\D_{q+3,\mu} \quad
(A_3)^{I+4}_{\mu}=-\frac{1}{2}A^{(5)I}_{\mu}+A^I_qA^q_{\mu}, \nonumber \\
&&(A_2)^{p+2}_{\mu}=\frac{1}{2}B\D_{p+3,\mu}-B_{pq}A^q_{\mu}+
\frac{1}{2}A^I_{p}A^{I+4}_{\mu},
\nonumber
\ee
which can be dualized on-shell as follows
\begin{eqnarray}
\nabla\times\overrightarrow{A_1}&=&\frac{1}{2}e^{2\p}G^{-1}
\left(\nabla u+(B+\frac{1}{2}AA^T)\nabla v+A\nabla s\right),
\nonumber                          \\
\nabla\times\overrightarrow{A_3}&=&\frac{1}{2}e^{2\p}
(\nabla s+A^T\nabla v)+A^T\nabla\times\overrightarrow{A_1},
\nonumber                            \\
\nabla\times\overrightarrow{A_2}&=&\frac{1}{2}e^{2\p}G\nabla v-
(B+\frac{1}{2}AA^T)\nabla\times\overrightarrow{A_1}+
A\nabla\times\overrightarrow{A_3}.
\end{eqnarray}
Here $u$ and $v$ are columns of dimension $2$ with conponents $u_1, u_2$
and $v_1, v_2$, respectively; and
the dimension of the column $s$ is $n$. So, the final system is defined by the quantities
$G$, $B$, $A$, $\p$, $u$, $v$ and $s$.  As it had been established in
\cite{hk1}--\cite{hk2}
%Refs.2--3
,
it is possible to introduce the matrix
Ernst potentials
\cite{e}
%^10
\be
\X=
\left(
\ba{cc}
-e^{-2\p}+v^TXv+v^TAs+\frac{1}{2}s^Ts&v^TX-u^T \cr
Xv+u+As&X
\ea
\right), \quad
\A=\left(
\ba{c}
s^T+v^TA \cr
A
\ea
\right),
\ee
where the $2 \times 2$ matrix potential $X=G+B+\frac{1}{2}AA^T$.
This pair of potentials allows us to express the
$3$-dimensional action (2) in a quasi--EM form
\cite {iw}--\cite {m}
:
%^{11--12}
\be
S^{(3)} = \int d^3 x \mid g\mid^{\frac{1}{2}}\,\{-R &+& {\rm Tr}[
\frac{1}{4}\left(\nabla \X - \nabla \A\A^T\right)\G^{-1}
\left(\nabla \X^T-\A\nabla \A^T\right)\G^{-1} \nonumber \\
&+& \frac{1}{2}\nabla \A^T\,\G^{-1}\nabla \A]\},
\ee
where $\G=\frac{1}{2}\left(\X+\X^T-\A\A^T\right)$.
This action leads to the following equations of motion
\be
&&\nabla^2\X -2(\nabla \X-\nabla \A A^T)(\X+\X^T-\A\A^T)^{-1}\nabla \X=0,
\nonumber
\\
&&\nabla^2\A -2(\nabla \X-\nabla \A A^T)(\X+\X^T-\A\A^T)^{-1}\nabla \A=0.
\ee
%%%%%%%%%%%%%%%%%%%%%%%%%%%%%%%%%%%%%%%%%%%%%%%%%%%%%%%%%%%%%%%%%%%%%%%%%%%%%%
%%%%%%%%%%%%%%%%%%%%%%%%%%%%%%%%%%%%%%%%%%%%%%%%%%%%%%%%%%%%%%%%%%%%%%
\section{BPS Saturated Dyon}
In this Sec. we obtain a class of extremal solutions for the
equations of motion (6) which generalize the IWP class of the EM theory
following the procedure indicated in
\cite{hk2}.
%Ref. 3
We consider a linear dependence between the potentials $\A$ and $\X$, and
require the matrix Ernst potentials to satisfy the asymptotic flatness
conditions $\X_{\infty} \rightarrow \Sigma$ and $\A_{\infty} \rightarrow 0$,
where $\Sigma=diag(-1,-1,1,1,...,1)$. This leads to the following relation
between the matrix Ernst potentials
\be
\A=(\Sigma - \X)b,
\ee
where $b$ is an arbitrary constant $3\times n$--matrix.
By substituting (7) into the action
(5) and setting the Lagrangian of the system to zero (it implies that
$R_{ij}=0$), we get the following condition to be satisfied
\be
bb^T=-\Sigma /2.
\ee
Indeed, {\it both} equations of motion (6) reduce to the Laplace equation
in Euclidean $3$--space
\be
\nabla ^2 [(\Sigma + \X)^{-1}]=0
\ee
which can be directly solved by the harmonic function (see
\cite{hk2}
%Ref. 3
for details)
\be
\frac{2}{\Sigma + \X}= \Sigma + \frac{M}{R},\quad {\rm where}\quad
\frac{1}{R}=Re\frac{1}{\sqrt{x^2+y^2+(z-i\a)^2}},
\ee
$M$ is a real $3$--dimensional arbitrary constant matrix and $\a$ is
a real constant. We choose $R$ in this way in order to deal with rotating
black hole solutions
\cite{bls}--\cite{sab}
%^{13--14}
(in this case we have a ring singularity). In order to obtain a real value of
the potential $\A$ (see Eqs. (7) and (8)) we can procede as follows:
Let us require just the first two rows of $b$ to be real (leaving the
remaining row imaginary), then we perform the matrix product (7) and set the
factors that multiply the imaginary components of $b$ to zero. It turns out
that this condition imposes the following restriction on the matrix
$M$
\be
M=\left(
\ba{ccc}
m_{11} & m_{12} & 0  \cr
m_{21} & m_{22} & 0  \cr
m_{31} & m_{32} & 0
\ea
\right)=\left(
\ba{cc}
M_{11} &  0  \cr

M_{21} &  0
\ea
\right)
\ee
leading to real solutions for the potential $\A$.

Now we begin to write down the explicit form of the single point--like
solution in terms of the five dimensional variables. In order to so, we
must calculate all vector three--fields using the dualization formulae (3).
Thus, after some algebraic manipulations we obtain
\be
\nabla \times \overrightarrow A^{(a)}=
m^{(a)}\nabla\left(\frac{1}{R}\right),
\ee
where $m^{(1)}=-(m_{12}-m_{21})/2$,
$m^{(2)}=m^{(4)}=m_{31}/2$, $m^{(3)}=-(m_{12}+m_{21})/2$  and
$m^{(4+n)}=(m_{11}b_{n1}+m_{12}b_{n2})$.

The relation between physical parameters and integration constants
becomes evident when we switch from Cartesian to oblate spheroidal
coordinates defined by
\be
x=\sqrt{\rho^2+\a^2}sin\theta cos\vp,\quad
y=\sqrt{\rho^2+\a^2}sin\theta sin\vp,\quad
z=\rho cos\theta,
\ee
In terms of these coordinates the $3$--interval reads
\be
ds^2_{3}=(\rho^2+\a^2cos^2\theta)(\rho^2+\a^2)^{-1}d\rho^2+
(\rho^2+\a^2cos^2\theta)d\theta^2+
(\rho^2+\a^2)sin^2\theta d\vp^2
\ee
and only the $A^{(a)}_{\vp}$ does not vanish\footnote{In fact we have
imposed the axial symmetry with respect to $z$.}:
\be
A^{(a)}_{\vp}=m^{(a)}cos\theta\frac{\rho^2+\a^2}
{\rho^2+\a^2cos^2\theta}=m^{(a)}\ep.
\ee
Studying the asymptotic behaviour of the $3$--fields we see that the
integration constants and the physical parameters of the theory are related
by
\be
G\sim
\left(
\ba{cc}
-\left(1+\frac{2m_{22}}{\rho}\right) &  \frac{m_{32}}{\rho}  \cr
\frac{m_{32}}{\rho} &  1
\ea
\right)=
\left(
\ba{cc}
-\left(1-\frac{2m}{\rho}\right) &  \frac{N_B}{\rho}  \cr
\frac{N_B}{\rho} &  1
\ea
\right), \quad B\sim \frac{m_{32}}{\rho}\sigma _2=
\frac{N_B}{\rho}\sigma _2,
\nonumber
\ee
\be
A=\left(
\ba{l}
A^{(n)}_t \cr
A^{(n)}_5
\ea
\right)
\sim
\left(
\ba{c}
2(m_{21}b_{1n}+m_{22}b_{2n})/\rho \cr
-2(m_{31}b_{1n}+m_{32}b_{2n})/\rho
\ea
\right)=
\left(
\ba{c}
Q^{(n)}_e/\rho \cr
Q^{(n)}_5/\rho
\ea
\right), \quad \p\sim -\frac{m_{11}}{\rho}=\frac{D}{\rho},
\nonumber
\ee
\be
u_1\sim \frac{m_{12}-m_{21}}{\rho}=\frac{N}{\rho}, \quad
v_1\sim \frac{m_{12}+m_{21}}{\rho}=\frac{Q_B}{\rho},
\nonumber
\ee
\be
u_2=v_2\sim \frac{m_{31}}{\rho}=\frac{N_5}{\rho}, \quad
s\sim 2\frac{m_{11}b_{n1}+m_{12}b_{n2}}{\rho}=
\frac{Q^{(n)}_m}{\rho},
\ee
where $m$ is the ADM mass, $D$, $N$,
$Q_B$ are the dilaton, NUT and axion charges, respectively; $N_B$ and $N_5$
are $5$--dimensional scalar charges, $Q^{(n)}_e$ and $Q^{(n)}_m$ are two
sets of $n$ electric and magnetic charges, and $Q^{(n)}_5$ are $n$ charges
that come from the extra dimension of the electromagnetic sector.
The extremality character of the found solutions makes these charges
to saturate the BPS bound
\be
4(D^2+m^2)+2(Q^2_B+N^2_u)+\sum_n (Q_5^{(n)})^2=
\sum_n (Q_e^{(n)})^2+\sum_n (Q_m^{(n)})^2+4(N^2_5+N^2_B),
\ee
this means that the attractive forces are precisely balanced
by the repulsive forces in the field configuration.

Let us count the number of independent parameters which parametrize the
physical charges of the solution. One of them is the rotational parameter
$\a$. The contribution of matrix $M$ is
equal to $6$. Matrix $b$ provides $2n-3$ independent parameters since only
its first two rows affect the solution and these rows are normalized and
orthogonal each other in view of Eq. (8). Thus we have $2n+4$ integration
constants which define the charges of the fields (in the case of arbitrary
$d$, the total number of independent parameters is $2(d+n)$).

The explicit form of the solution is given by the following relations
\be
ds^2=G_{MN}dx^{M}dx^{N}=G_{pq}\left(dx^{p+3}+\omega^{(p)} d\vp\right)
\left(dx^{q+3}+\omega^{(q)} d\vp\right)+
e^{2\p}g_{\mu\nu}dx^{\mu}dx^{\nu},
\ee
where the symmetric matrix $G_{pq}$ has the components
\be
(P^2+Q^2)^2G_{11}=
(Q^2-P^2)\left((\rho^2+2D\rho+\delta_0-\a^2cos^2\theta)
(\rho^2-\a^2cos^2\theta)\right.-
\nonumber
\ee
\be
\left.4(\rho+D)\rho\a^2cos^2\theta\right)-
4PQ\left((2\rho^2+3D\rho+\delta_0-2\a^2cos^2\theta)\rho-
D\a^2cos^2\theta\right)\a cos\theta, 
\nonumber
\ee
\be
(P^2+Q^2)^2G_{12}=
N_B\left((P^2-Q^2)(\rho^2-3\a^2cos^2\theta)\rho+
2PQ(3\rho^2-\a^2cos^2\theta)\a cos\theta\right)+
\nonumber
\ee
\be
(Q_BN_5+2DN_B)
\left((P^2-Q^2)(\rho^2-\a^2cos^2\theta)+
4PQ\rho\a cos\theta\right)+
\nonumber
\ee
\be
(\delta_1N_B-\delta_4N_5)
\left((P^2-Q^2)\rho+2PQ\a cos\theta\right),
\nonumber
\ee
\be
(P^2+Q^2)^2G_{22}=(P^2+Q^2)^2-
\delta_3
\left((P^2-Q^2)(\rho^2-\a^2cos^2\theta)+
4PQ\rho\a cos\theta\right)-
\nonumber
\ee
\be
2(Q_BN_5N_B+DN^2_B+mN^2_5)
\left((P^2-Q^2)\rho+2PQ\a cos\theta\right)-
\nonumber
\ee
\be
(\delta_1N^2_B+\delta_2N^2_5-2\delta_4N_5N_B)(P^2-Q^2),
\nonumber
\ee
the conformal multiplier is
\be
e^{2\p}=1+\frac{2D\rho}{\rho^2+\a^2cos^2\theta}+
\frac{\delta_0(\rho^2-\a^2cos^2\theta)}
{(\rho^2+\a^2cos^2\theta)^2}
\ee
and the components of the rotational vector are defined by
$\omega ^{(1)}=-N\ep$ and $\omega ^{(2)}=N_5\ep$.
Here we have introduced the following quantities
$P=\rho^2+(m+D)\rho+\Delta_1-\a^2cos^2\theta$,
$Q=(2\rho+(m+D))\a cos\theta$, $\delta_0=\delta_1-N^2_5$,
$\delta_1=D^2+\frac{1}{4}(Q_B-N)^2$,
$\delta_2=m^2+\frac{1}{4}(Q_B+N)^2$, $\delta_3=N^2_5+N^2_B$,
$\delta_4=\frac{1}{2}(m(N-Q_B)-D(N+Q_B))$ and
$\Delta_1=mD+\frac{1}{4}(N^2-Q^2_B)$.

The only non--vanishing components of the five--dimensional matter fields
are
\be
B=\frac{N_B(P\rho+Q\a cos\theta)-
\Delta_3P}
{P^2+Q^2}\sigma_2,
\nonumber
\ee
\be
A^{(n)}_t=
\frac{Q^{(n)}_e(P\rho+Q\a cos\theta)+
\left(DQ^{(n)}_e+\frac{1}{2}(Q_B-N)Q^{T(n)}_m\right)P}{P^2+Q^2},
\nonumber
\ee
\be
A^{(n)}_5=
\frac{Q^{(n)}_5(P\rho+Q\a cos\theta)-2
(\Delta_2b_{1n}-\Delta_3b_{2n})P}{P^2+Q^2},
\nonumber
\ee
\be
\p^{(5)}={\rm ln}
\left(\frac{P(\rho^2+2D\rho+\delta_0-\a^2cos^2\theta)+
2Q\a cos\theta (\rho+D)}{P^2+Q^2}
\right),
\nonumber
\ee
\be
B^{(5)}_{t\vp}=
-\frac{(N_BN_5+\frac{1}{2}Q^{(n)}_eQ^{(n)}_m)
(P\rho+Q\a cos\theta)-(\Delta_3 N_5+\Delta_1b_{2n}Q^{(n)}_m)P}
{P^2+Q^2}\epsilon -Q_B\ep,
\nonumber
\ee
\be
B^{(5)}_{5\vp}=
\frac{(\Delta_2b_{1n}Q^{(n)}_m+\Delta_3(N-Q^{(n)}_mb_{2n}))P-
(\frac{1}{2}Q^{(n)}_5Q^{(n)}_m+N_BN)
(P\rho+Q\a cos\theta)}{P^2+Q^2}\ep+N_5\ep
\nonumber
\ee
\be
A^{(5)I}_{\vp}=
\frac{(N_5Q^{T(n)}_5-NQ^{T(n)}_e)(P\rho+Q\a cos\theta)-
2(N_5\Delta_2b_{n1}-(N\Delta_1+N_5\Delta_3)b_{n2})P}{P^2+Q^2} \ep -
Q^{(n)}_m\ep,
\nonumber 
\ee
where $\Delta_2=mN_5+\frac {1}{2}N_B(Q_B-N)$, $\Delta_3=-[DN_B+\frac {1}{2}
N_5(Q_B+N)]$,
$b_{1n}=\frac {1}{4\Delta_1}[(Q_B+N)Q_e^{(n)}-2mQ_m^{T(n)}]$ and
$b_{2n}=-\frac {1}{4\Delta_1}[(Q_B-N)Q_m^{T(n)}+2DQ_e^{(n)}]$.
%%%%%%%%%%%%%%%%%%%%%%%%%%%%%%%%%%%%%%%%%%%%%%%%%%%%%%%%%%%%%%%%%%%%%%%%%%%%%%%%%
\section{Conclusions}
In this letter we have obtained a class of stationary extremal
solutions that generalize the IWP class of EM theory for the five--dimensional
heterotic string compactified to three dimensions on a two--torus.
These solutions are expressed in terms of $2n+4$ ($n\geq 3$ being the number
of Abelian vector fields) real parameters uniquely related to the physical
charges that saturate the BPS bound.

Among these solutions we identify rotating dyonic solutions with non--trivial
value of NUT parameter. If one requires the asymptotic flatness condition
to be satisfied in order to get a black hole configuration, one must set the
NUT parameter to zero; however, in this case the found solutions become
static.

%%%%%%%%%%%%%%%%%%%%%%%%%%%%%%%%%%%%%%%%%%%%%%%%%%%%%%%%%%%%%%%%%%%%%%%%%%%%%%
\section*{Acknowledgments}
We would like to thank our colleagues of NPI and JINR for
encouraging us during the performance of this letter. Authors
are grateful to N. Makhaldiani for helpful discussions. A.H.
was partially supported by CONACYT and SEP.
%%%%%%%%%%%%%%%%%%%%%%%%%%%%%%%%%%%%%%%%%%%%%%%%%%%%%%%%%%%%%%%%%%%%%%%%%%%%%%%

\end{document}